\documentclass[a4paper]{article}
\usepackage{graphicx}
\usepackage{amsmath}
\usepackage{amssymb}
\newcommand{\haak}[1]{\left(#1\right)}
\newcommand{\rhaak}[1]{\left [#1\right]}
\newcommand{\lhaak}[1]{\left | #1\right |}

\newcommand{\gem}[1]{\left\langle #1\right\rangle}

\newcommand{\geml}[1]{\left\langle #1\right.}
\newcommand{\gemr}[1]{\left. #1\right\rangle}
\newcommand{\haakl}[1]{\left(#1\right.}
\newcommand{\haakr}[1]{\left.#1\right)}
\newcommand{\rhaakl}[1]{\left[#1\right.}
\newcommand{\rhaakr}[1]{\left.#1\right]}
\newcommand{\lhaakl}[1]{\left |#1\right.}
\newcommand{\lhaakr}[1]{\left.#1\right |}
\newcommand{\ket}[1]{\lhaakl{\gemr{#1}}}
\newcommand{\bra}[1]{\lhaakr{\geml{#1}}}
\newcommand{\braket}[3]{\gem{#1\lhaak{#2}#3}}
\newcommand{\half}{\frac{1}{2}}
\newcommand{\kwart}{\frac{1}{4}}

\newcommand{\floor}[1]{\left\lfloor #1\right\rfloor}
\newcommand{\den}{\operatorname{d}}

\renewcommand{\imath}{{\text{i}}}

\begin{document}
\title{Exact expressions for correlations in the ground state of the dense 
$O(1)$ loop model} 
\author{S. Mitra$^{1}$, B. Nienhuis$^{1}$, J. de Gier$^{2}$ and M. T. 
Batchelor$^{3}$\\
$^{1}$Instituut voor Theoretische Fysica,\\
Universiteit van Amsterdam,\\
1018 XE Amsterdam, The Netherlands. 
\and
$^{2}$Department of Mathematics and Statistics,\\ The University of Melbourne, VIC 3010, Australia.
\and
$^{3}$Department of Theoretical Physics,\\ Research School of Physical Sciences \& Engineering\\ and Mathematical Sciences Institute,\\
Australian National University,\\ Canberra ACT 0200, Australia.}

\date{\today}
\maketitle
\begin{abstract}
Conjectures for analytical expressions for correlations in the dense  O$(1)$ 
loop model on semi infinite square lattices are given. We have obtained 
these results for four types of boundary conditions. Periodic and reflecting 
boundary conditions have been considered before. We give many new conjectures 
for these two cases and review some of the existing results. We 
also consider boundaries on which loops can end. We call such boundaries 
''open''. We have obtained expressions for correlations when both boundaries are 
open, and one is open and the other one is reflecting. Also, we formulate a 
conjecture relating the ground state of the model with open boundaries to Fully 
Packed Loop models on a finite square grid. We also review earlier obtained 
results about this relation for the three other types of boundary conditions. 
Finally, we construct a mapping between the ground state of the dense O$(1)$ 
loop model and the XXZ spin chain for the different types of boundary 
conditions.

\end{abstract}

\section{Introduction}
Recently Razumov and Stroganov \cite{Strog1,Raz1} made some remarkable
observations concerning the ground state of the antiferromagnetic XXZ
quantum chain of odd, finite length and with periodic boundary
conditions, and with anisotropy parameter $\Delta = -1/2$.
The ground state vector expressed in the standard basis of spin
configurations relative to the z-axis has all positive elements.
If these elements are normalized such that the smallest element is
unity, all elements turn out to have integer values. The most striking
observation is that some combinations of these integers are related to
the number of Alternating Sign Matrices (ASM) \cite{ASM1,ASM2}. These are
matrices of which the elements are equal to 0, 1 or -1, the non-zero elements alternate in sign and in each row and each column the elements add up to 1.

Since the first paper in the subject the relation between the XXZ chain
and the ASM has been extended considerably.
It was noted \cite{BdGN1} that the relation is more
generic if the XXZ Hamiltonian is reformulated in a different form, that
of a dense loop model.
This form is based on a well known equivalence \cite{baxkelwu} between the
$Q$-state Potts model at its critical point\cite{Wu},
the dense loop model and the 6-vertex model.
Of these two-dimensional statistical models the transfer matrix can be
taken to the limit of extreme spatial anisotropy where it takes a
simpler form which can be written as quantum Hamiltonian. This relates
the six-vertex model to the XXZ chain, and the equivalent dense loop
model to a Hamiltonian acting on configurations of arcs that pairwise
connect the sites of a chain. This loop Hamiltonian can be expressed
neatly in a graphical representation of the Temperley-Lieb
algebra\cite{PRGN,GNPR}.

It is in this formulation that the original authors \cite{Strog2}
discovered that the connection between the ground state of the loop
Hamiltonian and the ASM is considerably more detailed. This is based in
part on a simple
bijection between the class of ASM and the configurations of Fully
Packed Loop (FPL) models on a finite square grid with specific boundary
conditions.
It turns out that each element of the ground state vector of the loop
Hamiltonian is equal to the cardinality of a well defined subset of FPL
configurations. It turns out (see \cite{Strog3} and \cite{PRGN})
that different boundary conditions of the loop
Hamiltonian translate into different symmetry classes of FPL
configurations.
A review of these results has been presented by de Gier
\cite{degier02112852}.

In this paper we generalize these results to other boundary conditions
than have been considered so far. It is remarkable how robust the
results are under this type of variation. Also we give explicit
expressions for several classes of elements of the ground state vector
as well as for classes of correlation functions. We also
solve the reflection equation for the random cluster model, which
underlies the Hamiltonian for other than periodic boundary conditions.
Finally in order to make connection with the XXZ chain we give an explicit
transformation for vectors and operators in the loop representation to
the spin representation.

\section{The dense $O(n)$ loop model}
\begin{figure}[h]
\begin{center}
\includegraphics[width=0.5\textwidth]{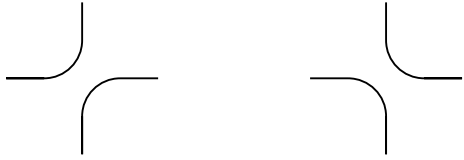}
\caption{The two vertices of the dense O(1) loop model}\label{fig:vrt}
\end{center}
\end{figure}

\begin{figure}[h]
\begin{center}
\includegraphics[width=0.5\textwidth]{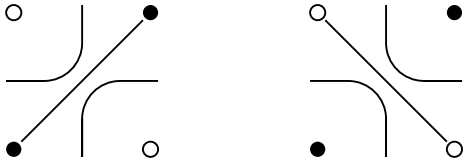}
\end{center}
\caption{The mapping of a loop configuration to a bond configuration of the corresponding bond percolation problem. A bond is either put on an edge of the square lattice formed by the $\bullet$ or on the dual edge 
orthogonal to it on the square lattice formed by the $\circ$.}\label{fig:perc}
\end{figure}

The states of the dense $O(n)$ loop model \cite{loop} are graphs consisting of 
non intersecting closed loops 
covering
all the edges of the lattice. Each vertex is visited
twice. Vertices can thus be in two states as shown in Figure \ref{fig:vrt}.
\newline The Boltzmann weight of a state consisting of $l$ loops is $n^{l}$. At $n=1$ all 
configurations
are equally likely. At this point the model can be
mapped to the bond percolation problem at criticality, the six vertex model and 
the XXZ-spin
chain at $\Delta=-1/2$ (see \cite{baxkelwu} and sec. \ref{mapping}). The corresponding bond percolation problem is defined on one of the sublattices of the dual lattice. 
The states of the
dense O$(1)$ loop model are in direct bijection with bond configurations of the 
bond percolation
problem, see Fig.\ \ref{fig:perc}. 
\setlength{\unitlength}{\textwidth}

In this article we will focus on the dense O$(1)$ loop model on an 
$L\times\infty$ lattice subjected to the
following boundary conditions:
\begin{itemize}
\item Periodic: The topology of the lattice is that of a cylinder.
\item Reflecting: At the boundaries the edges of even rows that end on the
boundary are connected to those of the odd row above it as indicated in
Fig.\ \ref{fig:refl}.
\begin{figure}[h]
\begin{center}
\includegraphics[width=0.25\textwidth]{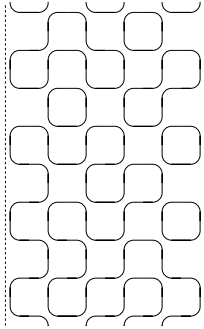}
\end{center}
\caption{Part of a typical configuration of the O$(1)$ loop model on a $6\times\infty$ strip with reflecting boundaries.}\label{fig:refl}
\end{figure}
\item mixed: Loops can end at the left boundary, while at the right boundary
reflecting boundary conditions are imposed.
\item open: Loops can end at both boundaries.
\end{itemize}

A horizontal cut between the vertices will intersect the loops at $L$ points. We 
define the connectivity state 
of the cut as the way these $L$ points are connected to each other or to the 
boundary by the loops via the 
half space below the cut. We will conjecture exact expressions for certain 
classes of connectivity states. 
Correlations can be defined as the probability that a subset of the points are 
connected in a prescribed way. 
We will also conjecture exact expressions for some correlations.
Connectivity states can be represented by a string of parentheses. If a point at 
position $i$
is connected to a point at position $j$, then this is represented by a 
parenthesis at position $i$ matching 
with a parenthesis at position $j$. If the point is connected to the left or 
right boundary, then that is indicated by a '')'', or ''('' respectively that doesn't match with any other parenthesis. If a point is not 
connected to any other point or to one of the boundaries then that is denoted by 
a ''$\lhaakl{}$''.

The expression $\haak{\ldots}_{k}$ shall stand for 
$\haak{\haak{\haak{\ldots}}}$, where $k$ delimiters have 
been 
opened and closed, and the dots symbolize an arbitrary well nested 
configuration.
For instance $\haak{\haak{\haak{}}}$ will be denoted as $\haak{}_{3}$, and
$\haak{\haak{\haak{}\haak{\haak{}}}}$ will be denoted as 
$\haak{\haak{}\haak{}_{2}}_{2}$.
We will omit subscripts equal to 1.
With a superscript we will denote a repeated concatenation of a structure with 
itself.
E.g. $\haak{}_{2}^{k}$ stands for a sequence of $k$ $\haak{}_{2}$: 
$\haak{}_{2}\haak{}_{2}\ldots$. Superscripts equal to 1 will be omitted. We will 
use the following notation 
for sequences of unpaired delimeters. By $\haakr{}_{k}$ we denote a sequence of 
$k$ ''$\haakr{}$'', and by 
$\haakl{}_{k}$ we denote a sequence of $k$ ''$\haakl{}$''. Note that we put the 
subscript always to the right 
of the delimeter. 
A configuration of the form $\haak{\ldots}_{m}$ will be referred to as a m-nest.
Note that an m-nest with no structure inside it spans $2m$ points.
An m-nest of this size will be referred to as a minimal m-nest. An m-nest 
spanning the entire system will be 
referred to as a maximal m-nest.

\section{The Hamiltonian and the transfer matrix}
When we consider a cut through an infinite strip or cylinder, the
probability distribution of specific connectivities at the cut
is precisely the distribution found in the eigenvector with the largest 
eigenvalue.

\begin{figure}[h]
\begin{center}
\includegraphics[angle=180,width=0.5\textwidth]{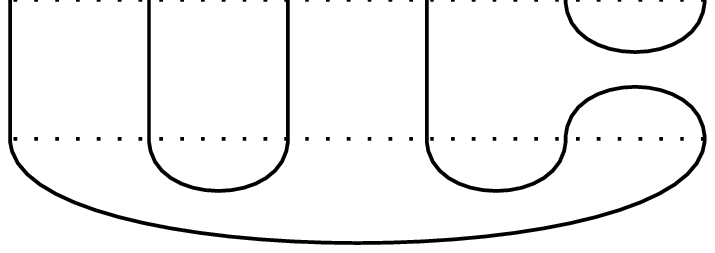}
\end{center}
\caption{The action of $e_{5}$ on $\haak{\haak{}\haak{}}$ resulting in $\haak{}_{2}\haak{}$}\label{fig:e1w1}
\end{figure}

As usual in integrable systems
the transfer matrix is a member of a family of commuting operators
parametrized by a spatial anisotropy. This anisotropy is introduced by
giving different weights to the two possible vertices of Fig.\ \ref{fig:vrt}.
In Fig.\ \ref{fig:vrtu} they are given the weight $\sin u$ and 
$\sin(\lambda - u)$, where $\lambda$ is related to the weight of a loop
$n = 2 \cos \lambda = (q + q^{-1})$. Eventually we will set $\lambda= \pi/3$, or equivalently $q ={\rm e}^{\pi{\rm i}/3}$, in order 
to have $n=1$, but the discussion in this section 
will be for general $\lambda$.
\setlength{\unitlength}{\textwidth}
\begin{figure}[h]
\begin{center}
\begin{picture}(1,0.3)
\put(0.15,0.05){\includegraphics[width=0.2\textwidth]{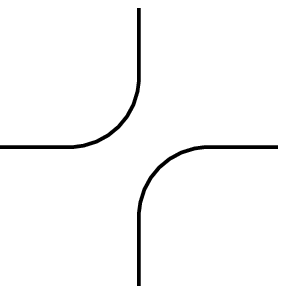}}
\put(0.65,0.05){\includegraphics[width=0.2\textwidth]{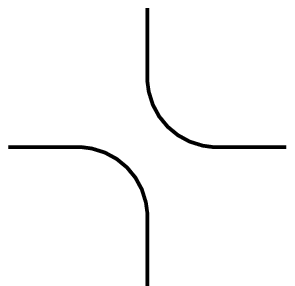}}
\put(0.25,0){\makebox(0,0){$\sin(u)$}}
\put(0.75,0){\makebox(0,0){$\sin(\lambda-u)$}}
\end{picture}
\caption{Boltzmann weights for a commuting set of transfer 
matrices}\label{fig:vrtu}
\end{center}
\end{figure}
\newline The Hamiltonian is now defined as the logarithmic derivative of the
transfer matrix
\begin{equation}
H = \frac{1}{T(0)} \left. \frac{d T(u)}{d u}\right|_{u=0}
\end{equation}
Up to an overal additive and a multiplicative constant the Hamiltonian
can be written as
\begin{equation}\label{hamilp}
H=\sum_{i=1}^{L}({1-e_{i}})
\end{equation}
where the operators $e_i$ generate the Temperley-Lieb algebra \cite{TempL71} 
\begin{eqnarray}
e_j^2 &=& (q + q^{-1}) \; e_j \nonumber\\
e_je_{j\pm1}e_j &=& e_j \label{eq:TLdef}\\
e_je_k &=& e_ke_j \quad |j-k| > 1, \nonumber
\end{eqnarray}
The action of the $e_i$ can be represented graphically as shown in
Fig.\ \ref{fig:ei}.
\setlength{\unitlength}{0.15\textwidth}
\begin{figure}[h]
\begin{scriptsize}
\begin{center}
\begin{picture}(3.6,1.6)
\put(0.8,0.5){\includegraphics[width=0.3\textwidth]{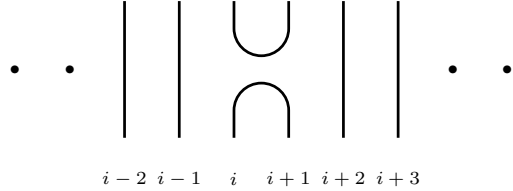}}
\put(0,1){\makebox(0,0){$\bullet$}}
\put(0.4,1){\makebox(0,0){$\bullet$}}
\put(3.2,1){\makebox(0,0){$\bullet$}}
\put(3.6,1){\makebox(0,0){$\bullet$}}
\put(0.8,0.2){\makebox(0,0){$i-2$}}
\put(1.2,0.2){\makebox(0,0){$i-1$}}
\put(1.6,0.2){\makebox(0,0){$i$}}
\put(2,0.2){\makebox(0,0){$i+1$}}
\put(2.4,0.2){\makebox(0,0){$i+2$}}
\put(2.8,0.2){\makebox(0,0){$i+3$}}
\end{picture}
\caption{The operators $e_{i}$}\label{fig:ei}
\end{center}
\end{scriptsize}
\end{figure}
\newline In this graphical notation, the component of the state vector is determined by
how the top row of line ends are connected by the figure below it. 
The action of an operator is visualised by placing the graph of the
operator above the graph of the state vector and keeping only the
information how the top row of lines are connected. See Fig.\ \ref{fig:e1w1} for 
an example.
Each closed loop gives an overall factor of $(q + q^{-1})$.
\setlength{\unitlength}{0.5\textwidth}

The Hamiltonian (\ref{hamilp}) is valid for the periodic system with an
additional algebraic relation among the $e_i$ \cite{levy,martin} to ensure that
loops winding around the cylinder are treated in the same way as
contractible loops.
When the loop model is placed on a strip rather than a cylinder, the
Hamiltonian is found as the logarithmic derivative of a family of
commuting double row 
transfer matrices. In the appendix we calculate the form of this double
row transfer matrix from the requirement that the boundary element
satisfies the reflection equation \cite{Sklyanin}.
There turns out to be a continuous family of boundary weights that
satisfies the reflection equation.
In the Hamiltonian limit they add to the TL algebra a left (right)
boundary element $h_0$ ($h_L$) which connects the leftmost (rightmost)
line to the boundary, as in Fig.\ \ref{fig:hdef}.
\setlength{\unitlength}{0.5\textwidth}
\begin{figure}[h]
\begin{center}
\begin{picture}(0.9,0.3)
\put(0.09,0){\includegraphics[height=0.15\textwidth]{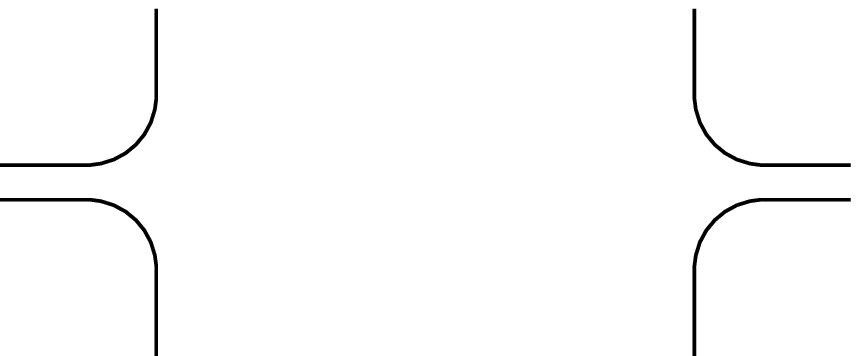}}
\put(0.01,0.15){\makebox(0,0){$h_{0}=$}}
\put(0.61,0.15){\makebox(0,0){$h_{L}=$}}
\end{picture}
\caption{Graphical definition of the operators $h_{0}$ and $h_{L}$}\label{fig:hdef}
\end{center}
\end{figure}
\newline These elements satisfy:
\begin{eqnarray}
h_0^2 = h_0 &\qquad& e_1 h_0 e_1 = e_1\nonumber \\
h_L^2 = h_L &\qquad& e_{L-1} h_L e_{L-1} = e_{L-1}\label{relhe}
\end{eqnarray}
The coefficient with which they appear in the Hamiltonian is arbitrary,
and we choose to consider the following cases. Closed or reflecting
boundary conditions:
\begin{equation}\label{hamilr}
H=\sum_{i=1}^{L-1}({1-e_{i}}),
\end{equation}
open boundary conditions
\begin{equation}\label{hamilo}
H= 1-h_0 + \sum_{i=1}^{L-1}({1-e_{i}}) + 1-h_L,
\end{equation}
and mixed boundary conditions
\begin{equation}\label{hamilm}
H= 1-h_0 + \sum_{i=1}^{L-1}({1-e_{i}}) .
\end{equation}
\section{Components of the ground state}
The lowest eigenvalue of each of the Hamiltonians defined in \eqref{hamilp}, \eqref{hamilr}, \eqref{hamilo} and \eqref{hamilm}
is zero for any value of $L$, and its corresponding eigenvector
possesses some intriguing properties. In this section we give conjectured analytical expressions for certain classes of components
of the eigenvector and also for certain classes of correlations. With the exception of a few simple cases, no 
proofs are available,
although some progress in this direction has been made recently \cite{janprf}.
We have computed the eigenvector up to $L=18$ for periodic boundaries, up to $L=16$ for closed boundaries,
up to $L=9$ for mixed boundaries, and up to $L=8$ for open boundaries.
We have normalized the eigenvector such that all components are integers with greatest common divisor 1.
Guessing analytical expressions for components amounts to guessing integer sequences from the first few terms.
There is no general practical method to do this, except if the sequence is given as
a product of a rational function of degree one or, a product of such products and so on, recursively. In such cases factorizing the numbers in the sequence will yield only small 
primes. If it is found that the first few integers in a sequence do indeed factor into small primes, then this is a strong indication that the sequence is of the above type. To find a formula of such a sequence one can transform the original sequence to one that is given as a rational function by applying a transformation $T$ to the sequence $a_{n}$, defined as
\begin{equation}
T\rhaak{a_{n}}=\frac{a_{n+1}}{a_{n}}
\end{equation}
often enough \cite{rate}. Given enough terms, this method will always succeed.
In practice one can often guess such sequences by simply inspecting the prime factorization of the terms in
the sequence, even if the above method fails due to a lack of terms.
For sequences that do not factor into small primes one can consult a database of integer sequences 
\cite{njas}.

The following functions enumerating various symmetry classes of 
alternating sign matrices will occur frequently in our expressions:
The number of $n\times n$ alternating sign matrices $A\haak{n}$ \cite{zeil,kup1}:
\begin{equation}
A\haak{n} =\prod_{j=0}^{n-1}\frac{\haak{3j + 1}!}{\haak{n + j}!}.
\end{equation}
The number of $2n\times 2n$ half turn invariant alternating sign
matrices \cite{rev}:
\begin{equation}
A_{\text{HT}}\haak{2n}=2\prod_{k=1}^{n-1}\frac{3\haak{3k + 2}!\haak{3k -
1}!k!\haak{k
-1}!}{4\haak{2k + 1}!^2\haak{2k - 1}!^2}.
\end{equation}
The conjectured number of $\haak{2 n-1}\times\haak{2 n-1}$ half turn
invariant alternating sign matrices \cite{rob}:
\begin{equation}
A_{\text{HT}}\haak{2n-1} = \prod_{j=1}^{n-1}\frac{4}{3}\frac{\haak{3j}!^2
j!^2}{\haak{2j}!^4}.
\end{equation}
The number of $\haak{2n+1}\times\haak{2n+1}$
vertically symmetric alternating sign matrices \cite{rev}:
\begin{equation}
A_{\text{V}}(2n+1) =\prod_{i=0}^{n-1} 
\frac{(3i+2)(6i+3)!(2i+1)!}{(4i+2)!(4i+3)!}.
\end{equation}
The number of cyclically symmetric transpose complement plane
partitions in a $\haak{2n}\times\haak{2n}\times\haak{2n}$ box \cite{macmahon} :
\begin{equation}
N_{8}\haak{2n}=\prod_{i=0}^{n-1}\frac{\haak{3i+1}\haak{6i}!\haak{2i}!}
{\haak{4i}!\haak{4i+1}!}.
\end{equation}
The conjectured number of $\haak{4 n\pm 1}\times\haak{4 n\pm 1}$ alternating sign matrices symmetric about 
both the horizontal and vertical axis \cite{rob}:
\begin{equation}
A_{\text{HV}}\haak{4 n\pm 1}=N_{8}\haak{2n}A_{\text{V}}\haak{2 n\pm 1}.
\end{equation} 
Some of the above combinatorial functions are special cases of the functions
$R\haak{n,p}$ and $Q\haak{n,p}$ defined below. We first define the auxiliary function:
\begin{equation}
\den\haak{n} = \prod_{k=1}^{n-1}\haak{2k -
1}!!2^{\floor{k/2}}.
\end{equation}
$Q\haak{n,p}$ and $R\haak{n,p}$ are defined as:
\begin{equation}
Q\haak{n, p} =\den\haak{n}^{-1}\rhaak{\prod_{k=1}^{\frac{n+1}{3}}\prod_{j=3
k-n}^{\frac{n +
3 -3k}{2}}\haak{p - j}}\rhaak{\displaystyle\prod_{k=1}^{\frac{n}{3}}\prod_{j=3k -
2n}^{-\frac{n + 3k}{2}}
\haak{2p +2j - 1}}
\end{equation}
\begin{equation}
R\haak{n, p} =\den\haak{n}^{-1}
\rhaak{\prod_{k=1}^{\frac{n + 1}{3}}\prod_{j=3k - 2n - 1}^{\frac{1 - n - 3
k}{2}}\haak{p + j}}
\rhaak{\displaystyle\prod_{k=1}^{\frac{n}{3}}\prod_{j=1 - n + 3k}^{\frac{2 + n -
3k}{2}}\haak{2p- 2j + 1}}
\end{equation}

The following relations hold:
\begin{equation}
\begin{split}
\haak{A\haak{n}}^{2}&=Q\haak{n,2n+1}\\
A\haak{n-1}A\haak{n-2}&=Q\haak{n,2n-2}\\
A_{\text{HT}}\haak{2n}&=Q\haak{n+1,2n+1}\\
A_{\text{HT}}\haak{2n+1}&= Q\haak{n+1,2n+2}\\
A_{\text{V}}\haak{2n+1}&=R\haak{n,2n+1}\\
N_{8}\haak{2n}&=R\haak{n,2n}\\
\end{split}
\end{equation}
\subsection{Periodic and reflecting boundary conditions}
The smallest components for even periodic and even reflecting systems are maximal
$L/2$ nests. For odd periodic and odd reflecting systems the smallest components
are maximal $L/2$ nests concatenated with an unpaired line. We normalize the eigenvector
by putting these components equal to one. We conjecture that with this normalization
all components are integers. For some components analytic expressions have been found. These components are 
listed in Table \ref{tabperref}.

\begin{table}
\begin{center}
\begin{tabular}{|c|c|c|c|}\hline
component&&periodic&reflecting\\\hline
$\haak{}_{L/2}$&&1&1\\\hline
$\haak{}_{L/2}\lhaakr{}$&&1&1\\\hline
$\haak{}^{L/2}$&&$A_{\text{HT}}\haak{L-1}$&$N_{8}\haak{L}$\\\hline
$\haak{}^{L/2}\lhaakr{}$&&$A\haak{\frac{L-1}{2}}^{2}$&$A_{\text{V}}\haak{L}$\\\hline
$\haak{\haak{}^{n}}_{\frac{L}{2}-n}$&&$Q\haak{n,L}$&$R\haak{n,L}$\\\hline
$\haak{\haak{}^{n}}_{\frac{L-1}{2}-n}\lhaakr{}$&&$Q\haak{n,L}$&$R\haak{n,L}$\\\hline
$\haak{\haak{}_{s}\haak{}_{t}}_{\frac{L}{2}-s-t}$&&see (\ref{pol})&see (\ref{ler}), (\ref{lor1}), 
(\ref{lor2})\\\hline
$\sum_{\ldots}\ldots$&e&$A_{\text{HT}}\haak{L}$&$A_{\text{V}}\haak{L+1}$\\\hline
$\sum_{\ldots}\ldots$&o&$A_{\text{HT}}\haak{L}$&$N_{8}\haak{L+1}$\\\hline
$\sum_{\ldots}\haak{\ldots}_{m}$&&$Q\haak{\frac{L}{2}-m,L+1}$&$R\haak{\frac{L}{2}-m,L+1}$\\\hline
$\sum_{\ldots}\haak{\ldots}_{m}\lhaakr{}$&&$Q\haak{\frac{L-1}{2}-m,L+1}$&$R\haak{\frac{L-1}{2}-m,L+1}$\\\hline
$\sum_{\ldots}\ldots\haak{}\ldots$&e&$\frac{3}{8}\frac{L^{2}}{L^{2}-1}A_{\text{HT}}\haak{L}$& see 
\eqref{refnstev}\\\hline
$\sum_{\ldots}\ldots\haak{}\ldots$&o&$\frac{3}{8}\frac{L^{2}-1}{L^{2}}A_{\text{HT}}\haak{L}$& see 
\eqref{refnstodd}\\\hline
$\sum_{\ldots}\ldots\haak{}_2\ldots$&e&see (\ref{nestev})&\\\hline
$\sum_{\ldots}\ldots\haak{}_2\ldots$&o&see (\ref{nestod})&\\\hline
\end{tabular}
\end{center}
\caption{Components of the eigenvector for periodic and reflecting boundary conditions. Where necessary, an e 
or o in the second column indicates if the system size is even or odd.}\label{tabperref}
\end{table}

In \cite{BdGN1} the following conjectures were made concerning the largest components and the sum of all 
components of periodic and reflecting systems.
For even periodic and reflecting systems the largest component is a sequence of
$L/2$ minimal 1-nests $\haak{}^{L/2}$. Its value is conjectured to be 
$A_{\text{HT}}\haak{L-1}$ for periodic systems and $N_{8}\haak{L}$ for reflecting systems. For odd periodic 
and reflecting systems the largest component is a sequence of $\haak{L-1}/2$ minimal 1-nests, concatenated with an unpaired line, $\haak{}^{\haak{L-1}/2}\lhaakl{}$. Its value is $A\haak{\haak{L-1}/2}^{2}$ for 
periodic systems and $A_{\text{V}}\haak{L}$ for reflecting systems.
The sum of all components of the eigenvector for periodic systems has been conjectured to be 
$A_{\text{HT}}\haak{L}$, for even reflecting systems this has been conjectured to be 
$A_{\text{V}}\haak{L+1}$, and for odd reflecting systems this has been conjectured to be $N_{8}\haak{L+1}$. 
We conjecture that two minimal nests side by side inside a maximal nest are given by certain (summations over) 
binomial determinants.

For periodic systems we conjecture that the components $\psi^{p}\haak{L,s,t}$ defined as
\begin{eqnarray}
\psi^{p}\haak{L,s,t} = &\haak{\haak{}_{s}\haak{}_{t}}_{\frac{L}{2}-s-t}&\text{ for even  }L\label{psipe}\\
\psi^{p}\haak{L,s,t} = &\haak{\haak{}_{s}\haak{}_{t}}_{\frac{L-1}{2}-s-t}\lhaakr{}&
\text{ for odd }L\label{psipo}
\end{eqnarray}
are given as the coefficient of $x^{s}$ of the polynomial: 
\begin{equation}\label{pol}
\det_{1\leq i,j \leq s+t}\rhaak{\binom{i+j-2+L-2s-2t}{i-1}+x\delta_{i,j}}
\end{equation}
For even reflecting systems this vector element, denoted as $\psi^{r}\haak{L,s,t}$
is conjectured to be \cite{rate}:
\begin{eqnarray}\label{deter}
\lefteqn{\psi^{r}\haak{L,s,t}=\det_{1\leq i,j \leq s} \rhaak{\binom{L+j-2i}{s+t-j}- 
\binom{L+j-2i}{s+t-j-2i+1}}=}&&\nonumber\\
&&\rhaakl{\prod_{j=1}^{s}\frac{\haak{j-1}!\haak{L-2s+2j-1}!\haak{L-2s-2t+3j-1}!}
{\haak{L-t-s+2j-1}!\haak{t+s-j}!\haak{L-2s-2t+2j-1}!}}
\times\nonumber\\
&&\rhaakr{\frac{\haak{L-s+t+2j-1}!}{\haak{L-2s+t+3j-1}!}}.\label{ler}
\end{eqnarray}

For odd reflecting systems we have less general results. The above formula for $\psi^{r}\haak{L,s,t}$ gives 
for $s+t=\haak{L-1}/2$ the vector element $\haak{}_{s}\lhaakr{}\haak{}_{t}$.
We conjecture that the vector element $\haak{\haak{}_{r}\haak{}}_{t}\lhaakr{}$ is given by:
\begin{equation}\label{lor1}
\binom{2r+2t+1}{r}-\binom{2r+2t+1}{r-3}.
\end{equation}
The vector element $\haak{}\haak{}_{r}\lhaakr{}$ is given by:
\begin{equation}\label{lor2}
\sum_{k=1}^{r+1}\frac{1}{k+1}\binom{2k}{k}.
\end{equation}

In the special cases of the vector elements $\psi\haak{L,1,1}$ and $\psi\haak{L,2,1}$, it is possible to prove 
the expressions that follow from the above formula using elementary manipulations involving the Hamiltonian.
We illustrate this for an even periodic system. According to \eqref{pol}, $\psi^{p}\haak{L,1,1}=L-1$. Let the 
connectivity states on which the loop Hamiltonian $H$ defined
in \eqref{hamilp} acts, be denoted as ket vectors. We can then write:
\begin{equation}
\braket{\phi}{H^{\dagger}}{\haak{}_{L/2}}=0
\end{equation}
where $\ket{\phi}$ is the eigenvector. Since
\begin{equation}
H^{\dagger}\ket{\haak{}_{L/2}}=-\ket{\haak{\haak{}\haak{}}_{L/2-2}}+\haak{L-1}\ket{\haak{}_{L/2}}
\end{equation}
it follows that the vector element $\psi^{p}\haak{L,1,1}$ has indeed the value $L-1$.

\subsection{Open and mixed boundary conditions}
For mixed and open boundary conditions the vector element consisting
of a sequence of L lines ending at left the boundary, $\haakr{}_{L}$
is the smallest component. In case of mixed boundary condition all components are conjectured to
be integers after normalizing the smallest component to 1. 
For open boundary conditions we define the coprime integers $V(L)$ and
$W(L)$  by,
\begin{equation}\label{vle}
\frac{V(L)}{W(L)} = \frac{A_\text{V}\haak{L+1}}{A_\text{V}\haak{L+3}},
\end{equation}
for even $L$, and for odd $L$,
\begin{equation}\label{vlo}
\frac{V(L)}{W(L)} = \frac{N_{8}\haak{L+1}}{N_{8}\haak{L+3}}.
\end{equation}
We conjecture that all components are integers when the smallest component is normalized to $V\haak{L}$. 
In Table \ref{tabmxop} we have listed the components for which we conjecture analytical 
expressions.
\begin{table}
\begin{small}
\begin{center}
\begin{tabular}{|c|c|c|}\hline
component&mixed&open\\\hline
$\haakr{}_{L}$&1&$V\haak{L}$, see (\ref{vle}) and (\ref{vlo})\\\hline
$\haak{}_{L/2}$&$A_{\text{V}}\haak{L+1}$&$V\haak{L}N_{8}\haak{L+2}$\\\hline
$\haak{}_{\haak{L-1}/2}\haakr{}$&$N_{8}\haak{L+1}$&$V\haak{L}A_{\text v}\haak{L+2}
$\\\hline
$\haak{}^{L/2}$&$A_{\text{V}}\haak{L+1}^{2}$&$V\haak{L}R\haak{\frac{L}{2},L+2}N_{8}\haak{L+2
}$\\\hline
$\haakr{}\haak{}^{\haak{L-1}/2}$&$A_{\text{V}}\haak{L}A_{\text{V}}\haak{L+2}$&$V\haak{L}R\haak{\frac{L-1}{2},L+2}
A_{\text{V}}\haak{L+2}$\\\hline

$\haakr{}\haak{}^{\haak{L-2}/2}\haakl{}$&& See \eqref{openlargest}\\\hline

$\haak{\haak{}^{n}}_{m-n}\haakr{}_{L-2m}$&$R\haak{n,L+1}R\haak{m+1,L+1}$&
$V\haak{L}R\haak{n,L+2}R\haak{m+1,L+2}$\\\hline
$\haakr{}\haak{}_{\haak{L-1}/2}$&$A_{\text{V}}\haak{L+2}$&\\\hline
$\haakr{}\haakr{}\haak{}_{\haak{L-2}/2}$&$A_{\text{V}}\haak{L+1}\frac{6}{L+4}\binom{L}{\haak{L-2}/2}$&\\\hline
$\haak{}\haak{}_{\haak{L-2}/2}$&$A_{\text{V}}\haak{L+1}\rhaak{\binom{L-1}{L/2}-\binom{L-1}{\haak{L+6}/2}}$&\\\hline
$\haak{}_{\haak{L-2}/2}\haak{}$&$A_{\text{V}}\haak{L+1}\sum_{k=1}^{L/2}\binom{2k}{k}$&\\\hline
$\haakr{}\haakr{}\haak{}^{\frac{L-2}{2}}$&$A_{\text{V}}\haak{L+1}^{2}$&\\\hline
$\sum_{\ldots}\ldots$&$A_{\text{HV}}\haak{2 L+3}$&$V\haak{L}A_{\text{HV}}\haak{2 L+5}$\\\hline
$\sum_{\ldots}\ldots\haakr{}_{n}$&see (\ref{pmx})&see (\ref{pop})\\\hline
\end{tabular}
\caption{Components of the eigenvector for mixed and open boundary conditions.}\label{tabmxop}
\end{center}
\end{small}
\end{table}
The largest component for even mixed systems is a sequence of
$L/2$ minimal 1-nests, $\haak{}^{L/2}$. The value of this component is 
given as $A_{\text{V}}\haak{L+1}^{2}$. For even open systems the largest component is $\haakr{}\haak{}^{\haak{L-2}/2}\haakl{}$ and its value is conjectured in \cite{pyatov} to be:
\begin{equation}\label{openlargest}
V\haak{L}\rhaak{\frac{A_{\text{V}}\haak{L+3}N_{8}\haak{L}R\haak{\frac{L}{2},L+2}}{A_{V}\haak{L+1}}-A_{\text{V}}\haak{L+1}R\haak{\frac{L}{2},L+3}}
.
\end{equation}
For an odd system the largest component is given by a line connected
to the left boundary placed to the left
of a sequence of minimal 1-nests, $\haakr{}\haak{}^{\haak{L-1}/2}$. For mixed systems these components are 
given as $A_{\text{V}}\haak{L}A_{\text{V}}\haak{L+2}$, and for open systems they are given as 
$V\haak{L}R\haak{\haak{L-1}/2,L+2}A_{\text{V}}\haak{L+2}$. The sum of all components is given by 
$A_{\text{HV}}\haak{2L+3}$ for mixed systems, and $V\haak{L}A_{\text{HV}}\haak{2L+5}$ for open systems.

\section{Correlations and exponents}
In this section we present a number of conjectured expressions for correlations. All conjectured correlations 
are  also listed in Tables \ref{tabperref} and \ref{tabmxop} where we give the corresponding sums of vector elements.
The probability for a minimal 1-nest in an even periodic system is:
\begin{equation}
\frac{3}{8}\frac{L^2}{L^2-1}.
\end{equation}
For odd periodic systems this probability is:
\begin{equation}
\frac{3}{8}\frac{L^2-1}{L^2}.
\end{equation}
In a reflecting system, the probability of a minimal 1-nest depends on its position. The average of this 
quantity over all positions for an even reflecting system is:
\begin{equation}\label{refnstev}
\frac{3 L^{2}+2 L+4}{\haak{L-1}\haak{8 L+4}}.
\end{equation}
 For an odd reflecting system this average is given as:
\begin{equation}\label{refnstodd}
\frac{3 L+5}{8 L+4}.
\end{equation}
The probability for a minimal 2-nest in an even periodic system is:
\begin{equation}\label{nestev}
\frac{\haak{L-2}\haak{59 L^5+118 L^4-44 L^3-88L^2+5760L-28800}}
{2^{10}\haak{L^2-9}\haak{L^2-1}^2}.
\end{equation}
And for an odd periodic system it is:
\begin{equation}\label{nestod}
\frac{\haak{L-3}\haak{59 L^6+531L^5+1460L^4-750L^3+3949L^2-20001L-1890}}
{2^{10}\haak{L-2}L^3\haak{L+2}^2\haak{L+4}}.
\end{equation}

Because for many correlations the asymptotic large $L$ dependence is
algebraic with a non-trivial exponent, we give this behavior using
the symbol $\propto$. 
The probability for a maximal m-nest in even periodic systems is:
\begin{equation}
\frac{Q\haak{\frac{L}{2}-m,L+1}}{A_{\text{HT}}\haak{L}}\propto L^{-\haak{1+m}\haak{1+2m}/3}.
\end{equation}
And for odd periodic systems this probability becomes:
\begin{equation}
\frac{Q\haak{\frac{L-1}{2}-m,L+1}}{A_{\text{HT}}\haak{L}}\propto L^{-\haak{1+m}\haak{3+2m}/3}.
\end{equation}
The probability for a maximal m-nest in an even reflecting system is:
\begin{equation}
\frac{R\haak{\frac{L}{2}-m,L+1}}{A_{\text{V}}\haak{L+1}}\propto L^{-\frac{2}{3}m\haak{m+1}}.
\end{equation}
For an odd reflecting system this probability is:
\begin{equation}
\frac{R\haak{\frac{L-1}{2}-m,L+1}}{N_{8}\haak{L+1}}\propto L^{-\frac{2}{3}\haak{m+1}^{2}}.
\end{equation}

In an even periodic system the probability $P\haak{L,n}$ that $n$ consecutive points are disconnected from each other is given as:
\begin{equation}
P\haak{L,n}=\frac{S\haak{L,n}}{S\haak{2n,n} A\haak{n}}
\end{equation}
where $S\haak{L,n}$ for even $n$ is given as:
\begin{equation}
\frac{\prod_{p=1}^{n/2}\prod_{k=p}^{2p-1}\haak{L^{2}-4 k^{2}}}{\prod_{p=0}^{n/2-1}\haak{L^{2}-
\haak{2 p+1}^{2}}^{n/2-p}}
\end{equation}
while for odd $n$ it is
\begin{equation}
\frac{\prod_{p=2}^{\haak{n+1}/2}\prod_{k=p}^{2p-2}\haak{L^{2}-4 
k^{2}}}{\prod_{p=0}^{\haak{n-3}/2-1}\haak{L^{2}-
\haak{2 p+1}^{2}}^{\haak{n-1}/2-p}}.
\end{equation}
The function $f\haak{n}\equiv\lim_{L\rightarrow\infty}P\haak{L,n}$ has the following asymptotic
behavior for large $n$:
\begin{equation}
f\haak{n}\propto 4^{-n\haak{3n+2}/4}\haak{3\sqrt{3}}^{n\haak{n+1}/2}n^{7/72}.
\end{equation}

For mixed and open boundary conditions we have obtained the probability that at least the $n$
rightmost sites are connected to the left boundary. For mixed boundary conditions, we conjecture that the sum 
of the corresponding vectorelements $P_{\text{mx}}\haak{L,n}$ is given as:
\begin{equation}\label{pmx}
P_{\text{mx}}\haak{L,n}=R\haak{\floor{\frac{L-n+1}{2}},L+1}
R\haak{\floor{\frac{L-n+2}{2}},L+2}.
\end{equation}
The probability decays as:
\begin{equation}
\frac{P_{\text{mx}}\haak{L,n}}{P_{\text{mx}}\haak{L,0}}\propto L^{-n\haak{1+n}/3}.
\end{equation}
For open boundary conditions, the sum of vector elements in which at
least the $n$ righmost lines are connected to the left boundary, is
conjectured to be:
\begin{equation}\label{pop}
P_{\text{op}}\haak{L,n}=V\haak{L}R\haak{\floor{\frac{L-n+1}{2}},L+2}
R\haak{\floor{\frac{L-n+2}{2}},L+3}.
\end{equation}
The normalized probability decays as:
\begin{equation}
\frac{P_{\text{op}}\haak{L,n}}{P_{\text{op}}\haak{L,0}}\propto L^{-n\haak{3+n}/3}.
\end{equation}

\section{Fully packed loop diagrams}
As we have indicated above, the sum of the components of the
eigenvector of the transfer matrix of the dense O(1) loop model is
conjectured to be the number of ASMs of a certain symmetry class. The
specific symmetry class is determined by the boundary condition of the
O(1) loop model. In this section this connection will be specified
further.

\begin{figure}
\begin{center}
\includegraphics[width=0.5\textwidth]{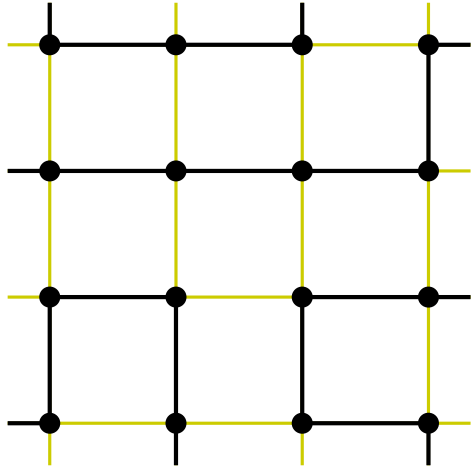}
\end{center}
\caption{An FPL diagram.}\label{fpl}
\end{figure}

It is known (see e.g.\ \cite{elk}) that ASMs are in bijection
with certain classes of fully packed loop (FPL) diagrams on square
grids. A grid is a rectangular section of the square lattice, of which
all vertices are  incident on four edges, and on each edge one or two
vertices are incident. The edges on which only one vertex is incident
are called external or boundary edges. An FPL diagram on a grid is a
collection of lattice paths such that each vertex is visited once by one
of the paths. Each of the paths is cyclic or open. In the latter case it
runs from an external edge to another external edge. For later reference
we number the external edges anticlockwise starting with the uppermost
horizontal edge on the left side of the grid. Fig.\ \ref{fpl} shows an
example of an FPL diagram on a square grid. 

The mapping between FPL diagrams and ASMs is symmetry preserving. The
conjectured equality between the sum of the components of the
eigenvector and the number of ASMs of certain symmetry thus extends to
the number of FPL diagrams of the same symmetry. Consider FPL diagrams
on a square grid in which the even numbered external edges are visited
by the paths. By the connectivity of the FPL diagram we denote the way
in which these external edges are pairwise connected by the paths. 
Razumov e.a.\ \cite{Strog2} conjectured that the components of the
groundstate eigenvector of the O(1) loop model with periodic identified
boundaries are equal to the number of $L/2\times L/2$ FPL diagrams with the
corresponding connectivity. This connection was later generalized to
other boundary conditions: periodic unidentified boundaries
\cite{Strog3} to the class of half-turn invariant FPL diagrams, 
reflecting boundary conditions \cite{PRGN} to vertically symmetric FPL
diagrams.
Here we generalize it to open and mixed boundary conditions. The latter
was already reported in \cite{degier02112852}.

\begin{figure}
\begin{center}
\includegraphics[width=0.5\textwidth]{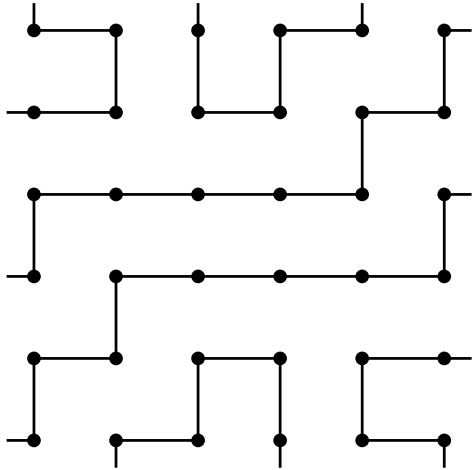}
\end{center}
\caption{Half-turn invariant FPL diagram corresponding to the vector element $\haak{\haak{}\haak{}}$.}\label{fplht}
\end{figure}

The case of periodic unidentified boundaries maps into the class of
half-turn invariant FPL diagrams on an $L\times L$ grid, 
see Fig.\ \ref{fplht} for an example. Again only even numbered external
edges are visited. They represent a row of vertical edges of the O(1)
loop model. 

\begin{figure}
\begin{center}
\includegraphics[width=0.5\textwidth]{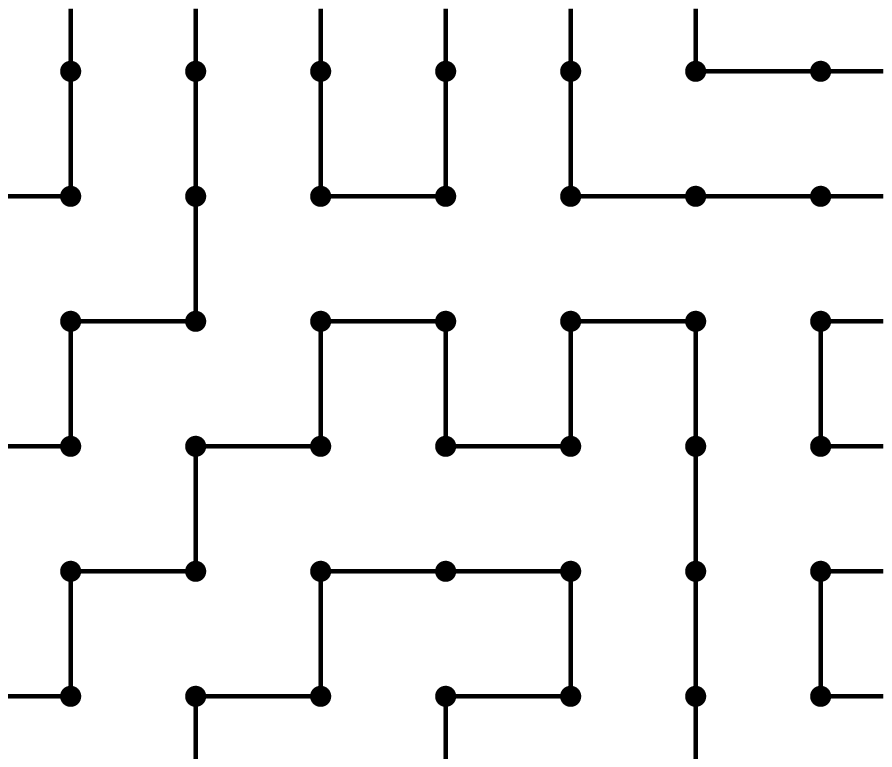}
\end{center}
\caption{An FPL diagram corresponding to the vector element 
$||(())$.}\label{fplo}
\end{figure}

\begin{itemize}
\item {\it Reflecting boundaries.}
The eigenvector components of the
O(1) loop model on a strip with even size $L$ and reflecting boundary
conditions map onto a vertically symmetric FPL diagram on a
$(L+1)\times(L+1)$ grid. This symmetry indicates that the FPL
configurations are completely determined by the $L/2 \times (L+1)$
rectangle. Let the long sides of the rectangle be horizontal, then
external edges of the top side are not visited by the paths. 
Of the remaining three sides the even-numbered 
edges are visited. The number of these FPL diagrams with a
given connectivity is conjectured to be the component of eigenvector
with that connectivity.
Because in the $L/2\times (L+1)$ geometry the path configuration at the
left and right sides is completely fixed, it is sufficient to specify only a
$L/2 \times (L-1)$ rectangle, with now the odd-numbered external edges
visited.

A very similar conjecture has been given 
for an odd sized system with reflecting boundaries.
Here the rectangle is $(L-1)/2 \times L$, again oriented with the long
sides horizontal. Of the top side (of size $L$)
precisely one of the external edges is
visited, and of the other sides the odd-numbered edges. 
One of the latter is connected to the exceptional top side edge, and
represents the unpaired site of the O(1) model.

\item {\it Mixed boundary conditions.} 
When one boundary of the O(1) strip is reflecting and the other is open,
the sum of the ground state vector elements we conjecture to be equal to
the number of horizontally and vertically symmetric $(2L+3)\times(2L+3)$
FPL diagrams (see also \cite{degier02112852}). Again the symmetry
prescribes that one quadrant of the grid completely determines the
configuration. An example is shown in Fig.\ \ref{fplmx}. Also the
configuration at the boundary sites is determined by the requirement
that all sites be visited by the paths.  As a result the eigenvector
corresponds to the FPL diagrams on an $L\times L$ grid. Of the left and
bottom side the even-numbered external edges are visited, and all of the
top side. The left and bottom side represent the sites of the O(1)
vector. Those of the top side represent the open boundary. The
connectivity is defined by the way the left and bottom edges are
mutually connected, and by which of them are connected to any edge of
the top side. We conjecture the number of FPL diagrams with a given
connectivity to be equal to the eigenvector element with the
corresponding connectivity.

\begin{figure}
\begin{center}
\includegraphics[width=0.5\textwidth]{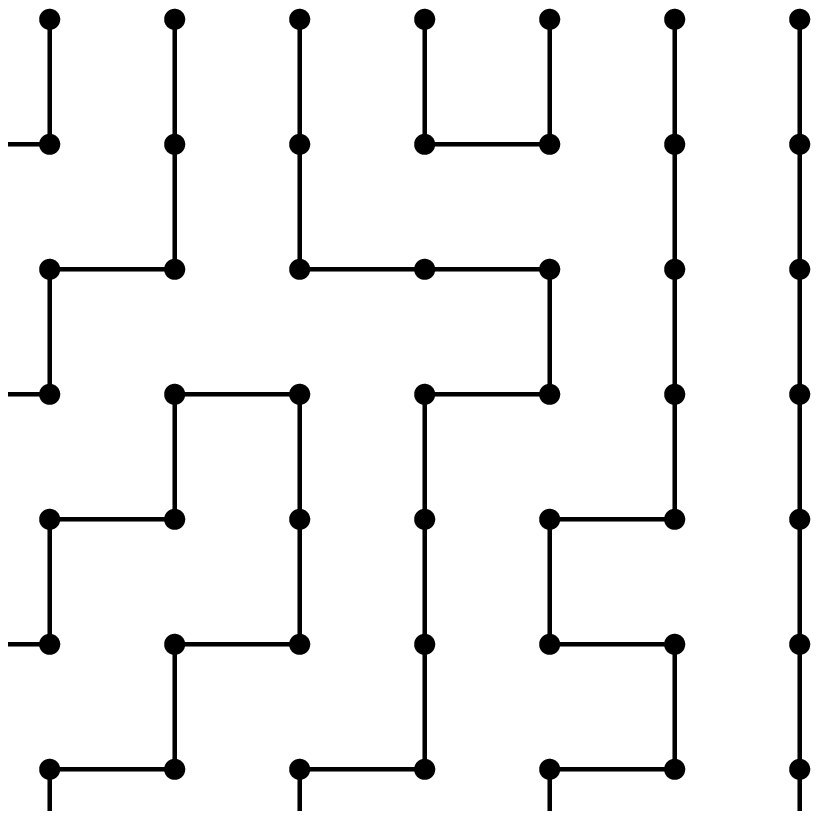}
\end{center}
\caption{An FPL diagram corresponding to the vector element 
$\haakr{}\haak{}\haakr{}\haakr{}$.}\label{fplmx}
\end{figure}

\item {\it Open boundaries.}
When the O(1) loop model has two open boundaries, one may or may not
distinguish these boundaries in defining the connectivity. In the data
presented above we do make the distinction if a site is connected to the
right or the left boundary. For this case we do not have conjectures to
connect the individual vector components to FPL counts. 
We can, however, make the connection when we identify the two
boundaries in the definition of the connectivity.
For odd system size $L$ the numbers are derived from the class of
vertically symmetric FPL diagrams of size $(L+2)\times(L+2)$. These are
completely specified by the configuration on a $L\times(L+1)/2$
rectangle. The external edges of the top side, of size $L$, are all
visited by the paths. These represent the boundary. Of the remaining
sides the odd-numbered external edges are visited, and the connectivity
specifies how these are connected to each other. The components of the
eigenvector of the O(1) model is conjectured to be the number of FPL
diagrams with the corresponding connectivity.

For even system size $L$ and identified open boundaries we have the
following conjecture. Consider FPL digrams on a $L\times(L+1)$ grid.
Two adjacent sides represent the boundary. Of these all external edges 
except $(L+1)$st one on the long side are visited by the paths. 
Of the other two sides the even-numbered edges are visited, and
represent the row of edges of the O(1) model.
The connectivity specifies
which of these edges are pairwise connected by the paths and which are
connected to the boundary. An example is shown in 
Fig.\ \ref{fplo}. The number of FPL
diagrams with a given connectivity is conjectured to be equal to the
eigenvector component with the same connectivity. In this case, the
smallest element is not equal to one, but given by 
$A_{\text{V}}\haak{L+1}$.  
\end{itemize}

\section{Interpretation as families of nonintersecting lattice
paths}\label{secpath}
In this section we will show that the conjectured expressions for the components 
$\psi^{p}\haak{L,s,t}$ and $\psi^{r}\haak{L,s,t}$, see \eqref{psipe}, 
\eqref{psipo} and \eqref{ler}, enumerate families of nonintersecting lattice 
paths on finite square lattices. The lattice paths consist of unit steps from 
points to one of their nearest neighbors. The path may only move in two mutually orthogonal
directions, e.g.\ in the positive $x$ or negative $y$ 
direction.

According to Gessel and Viennot's theorem on the enumeration of nonintersecting
lattice paths \cite{gessel}:

Let $A_{1},A_{2}\ldots A_{n}$ and $E_{1},E_{2}\ldots E_{n}$ be lattice
points
such that for $i_{1}<i_{2}$ and $j_{1}<j_{2}$ any lattice path from
$A_{i_{1}}$ to
$E_{j_{2}}$ meets any lattice path from $A_{i_{2}}$ to $E_{j_{1}}$.
Then the number of families
$\haak{P_{1},P_{2},\ldots,P_{n}}$ of nonintersecting lattice paths, where
$P_{i}$ runs
from $A_{i}$ to $E_{i}$ is given by the determinant:
\begin{equation}
\det_{1\leq i,j\leq n} Z\haak{A_{i}\rightarrow E_{j}}
\end{equation}
where $Z\haak{A_{i}\rightarrow E_{j}}$ denotes the number of all lattice
paths from
$A_{i}$ to $E_{j}$

We will now show that the conjectured expression for $\psi^{p}\haak{L,s,t}$ 
enumerates
the total number of $s$ nonintersecting lattice paths on an
$\haak{s+t}\times\haak{L-s-t}$ square lattice
subjected to the constraints:
\begin{enumerate}
\item A path is required to start at a point $\haak{0,k}$ with $0\leq
k\leq s+t-1$.
\item A path starting at $\haak{0,k}$ is required to end at the point
$\haak{k+L-2s-2t,0}$.
\end{enumerate}
See Fig.\ \ref{fig:path} for an example.
\setlength{\unitlength}{0.5\textwidth}
\begin{figure}[h]
\begin{center}
\begin{picture}(1,0.8)(-0.08,-0.08)
\put(0,0){\makebox(0,0){$\bullet$}}
\put(0.2,0){\makebox(0,0){$\bullet$}}
\put(0.4,0){\makebox(0,0){$\bullet$}}
\put(0.6,0){\makebox(0,0){$\bullet$}}
\put(0.8,0){\makebox(0,0){$\bullet$}}
\put(1,0){\makebox(0,0){$\bullet$}}

\put(0,0.2){\makebox(0,0){$\bullet$}}
\put(0.2,0.2){\makebox(0,0){$\bullet$}}
\put(0.4,0.2){\makebox(0,0){$\bullet$}}
\put(0.6,0.2){\makebox(0,0){$\bullet$}}
\put(0.8,0.2){\makebox(0,0){$\bullet$}}
\put(1,0.2){\makebox(0,0){$\bullet$}}

\put(0,0.4){\makebox(0,0){$\bullet$}}
\put(0.2,0.4){\makebox(0,0){$\bullet$}}
\put(0.4,0.4){\makebox(0,0){$\bullet$}}
\put(0.6,0.4){\makebox(0,0){$\bullet$}}
\put(0.8,0.4){\makebox(0,0){$\bullet$}}
\put(1,0.4){\makebox(0,0){$\bullet$}}

\put(0,0.6){\makebox(0,0){$\bullet$}}
\put(0.2,0.6){\makebox(0,0){$\bullet$}}
\put(0.4,0.6){\makebox(0,0){$\bullet$}}
\put(0.6,0.6){\makebox(0,0){$\bullet$}}
\put(0.8,0.6){\makebox(0,0){$\bullet$}}
\put(1,0.6){\makebox(0,0){$\bullet$}}

\put(-0.05,0){\makebox(0,0){$A_{4}$}}
\put(-0.05,0.2){\makebox(0,0){$A_{3}$}}
\put(-0.05,0.4){\makebox(0,0){$A_{2}$}}
\put(-0.05,0.6){\makebox(0,0){$A_{1}$}}

\put(0.4,-0.05){\makebox(0,0){$E_{4}$}}
\put(0.6,-0.05){\makebox(0,0){$E_{3}$}}
\put(0.8,-0.05){\makebox(0,0){$E_{2}$}}
\put(1,-0.05){\makebox(0,0){$E_{1}$}}

\put(0,0.6){\line(1,0){0.4}}
\put(0.4,0.6){\line(0,-1){0.2}}
\put(0.4,0.4){\line(1,0){0.4}}
\put(0.8,0.4){\line(0,-1){0.2}}
\put(0.8,0.2){\line(1,0){0.2}}
\put(1,0.2){\line(0,-1){0.2}}

\put(0,0.4){\line(1,0){0.2}}
\put(0.2,0.4){\line(0,-1){0.2}}
\put(0.2,0.2){\line(1,0){0.2}}
\put(0.4,0.2){\line(0,-1){0.2}}
\put(0.4,0){\line(1,0){0.4}}
\end{picture}
\caption{A contribution to $\psi^{p}\haak{L=10,s=2,t=2}$.
Paths starting at $A_{i}$ have to end at $E_{i}$.}\label{fig:path}
\end{center}
\end{figure}
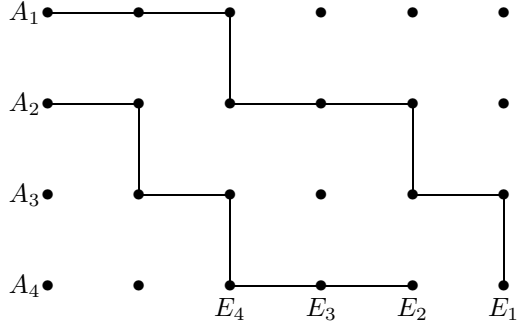
\newline Using Gessel and Viennot's theorem we can write the total number of
families of nonintersecting
lattice paths subjected to the above constraints as:
\begin{equation}
\sum_{0\leq p_{1}<p_{2}<\ldots <p_{s-1}<p_{s}\leq s+t-1}\det_{1\leq i,j\leq
s}\binom{p_{j}+p_{i}+2p}{p_{i}}.
\end{equation}
Note that this is precisely the sum of all s-th order principal minors of
the matrix $M_{i,j}=\binom{i+j+2p}{i}$, where $0\leq i,j\leq s+t-1$. In general, 
the sum of s-th order principal minors of an arbitrary $n\times n$ matrix will 
yield
the absolute value of the coefficient of $x^{n-s}$ of the characteristic 
polynomial
of the matrix. It thus follows that the total number of nonintersecting lattice
paths is given by the absolute value of the coefficient of $x^{t}$ of the
characteristic polynomial of the matrix $M_{i,j}$.

The conjectured expression for $\psi^{r}\haak{L,s,t}$ \eqref{ler} gives the 
number of
$s$ nonintersecting lattice paths subjected to the constraints:
\begin{enumerate}
\item The $k$-th path $\haak{1\leq k\leq s}$ starts at the point
$\haak{2k-1,0}$ and ends at the point
$\haak{L-s-t+2k-1,s+t-k}$.
\item every path stays below the diagonal line $\haak{x,x}$.
\end{enumerate}
See Fig.\ \ref{fig:pathd} for an example.
\setlength{\unitlength}{0.06\textwidth}
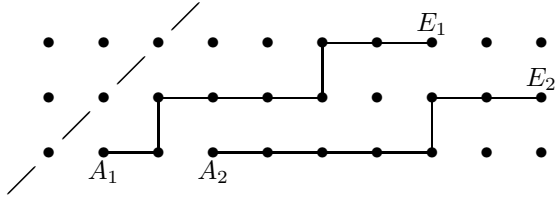
\begin{figure}[h]
\begin{center}
\begin{picture}(11,6)(-1,-1)
\put(0,0){\makebox(0,0){$\bullet$}}
\put(1,0){\makebox(0,0){$\bullet$}}
\put(2,0){\makebox(0,0){$\bullet$}}
\put(3,0){\makebox(0,0){$\bullet$}}
\put(4,0){\makebox(0,0){$\bullet$}}
\put(5,0){\makebox(0,0){$\bullet$}}
\put(6,0){\makebox(0,0){$\bullet$}}
\put(7,0){\makebox(0,0){$\bullet$}}
\put(8,0){\makebox(0,0){$\bullet$}}
\put(9,0){\makebox(0,0){$\bullet$}}

\put(0,1){\makebox(0,0){$\bullet$}}
\put(1,1){\makebox(0,0){$\bullet$}}
\put(2,1){\makebox(0,0){$\bullet$}}
\put(3,1){\makebox(0,0){$\bullet$}}
\put(4,1){\makebox(0,0){$\bullet$}}
\put(5,1){\makebox(0,0){$\bullet$}}
\put(6,1){\makebox(0,0){$\bullet$}}
\put(7,1){\makebox(0,0){$\bullet$}}
\put(8,1){\makebox(0,0){$\bullet$}}
\put(9,1){\makebox(0,0){$\bullet$}}

\put(0,2){\makebox(0,0){$\bullet$}}
\put(1,2){\makebox(0,0){$\bullet$}}
\put(2,2){\makebox(0,0){$\bullet$}}
\put(3,2){\makebox(0,0){$\bullet$}}
\put(4,2){\makebox(0,0){$\bullet$}}
\put(5,2){\makebox(0,0){$\bullet$}}
\put(6,2){\makebox(0,0){$\bullet$}}
\put(7,2){\makebox(0,0){$\bullet$}}
\put(8,2){\makebox(0,0){$\bullet$}}
\put(9,2){\makebox(0,0){$\bullet$}}

\put(-0.75,-0.75){\line(1,1){0.5}}
\put(0.25,0.25){\line(1,1){0.5}}
\put(1.25,1.25){\line(1,1){0.5}}
\put(2.25,2.25){\line(1,1){0.5}}

\put(1,0){\line(1,0){1}}
\put(2,0){\line(0,1){1}}
\put(2,1){\line(1,0){3}}
\put(5,1){\line(0,1){1}}
\put(5,2){\line(1,0){2}}

\put(3,0){\line(1,0){4}}
\put(7,0){\line(0,1){1}}
\put(7,1){\line(1,0){2}}

\put(1,-0.35){\makebox(0,0){$A_{1}$}}
\put(3,-0.35){\makebox(0,0){$A_{2}$}}
\put(7,2.35){\makebox(0,0){$E_{1}$}}
\put(9,1.35){\makebox(0,0){$E_{2}$}}

\end{picture}
\caption{A contribution to $\psi^{r}\haak{L=10,s=2,t=2}$.
Paths starting at $A_{i}$ have to end at $E_{i}$ and are not allowed to touch the dashed line.}\label{fig:pathd}
\end{center}
\end{figure}
\newline The proof that the number of families of nonintersecting lattice paths
corresponds to the
expression (\ref{ler}) is again a straightforward application of Gessel and
Viennot's theorem.
From the reflection principle it follows that the total number of paths from 
the point $\haak{a,0}$ to the point
$\haak{m,n}$, with $m\geq n$
that stay below the diagonal line $\haak{x,x}$ is given by (see e.g.\ 
\cite{mohanty})
\begin{equation}\label{bal}
B\haak{a,m,n}=\binom{m+n-a}{n} - \binom{m+n-a}{m}.
\end{equation}
It thus follows that the total number of families of nonintersecting lattice
paths is given
as the determinant:
\begin{equation}
\det_{1\leq i,j\leq s} B\haak{2 i-1,L+r-s-2t+2j-1,r+s-j}.
\end{equation}
Inserting (\ref{bal}) in this determinant yields (\ref{ler}).

\section{Mapping Loop Model $\rightarrow$ ground state of XXZ Model}\label{mapping}
In this section we derive the mapping from the dense O(1) loop model to the ground
state of the XXZ model for all the boundary conditions considered in
this article. Proceeding in a similar way as in \cite{baxkelwu}, we first add an extra degree of freedom to the loops of the dense O(1) model in the form of an orientation. We assign a Boltzmann weight of $q^{1/4}$ for each left turn and $q^{-1/4}$ for each 
right turn. The loopweight thus becomes $q+q^{-1}$. The choice $q=\exp\haak{\frac{\imath\pi}{3}}$ ensures that this is unity.

In this model a horizontal cut between the vertices is now intersected by arrows. This can be interpreted as a state of a spin chain of length $L$. The Hamiltonian of the loop model \eqref{hamilp} defines a Hamiltonian acting on the spin chain as follows:
The operator $e_{i}$ (see Fig.\ \ref{fig:ei}) can be transformed to a spin operator
by orienting the two half loops, assigning the correct Boltzmann weights
and summing over the orientations.
We obtain:
\begin{equation}\label{hamder}
e_{i}=\haak{q^{-\half}\ket{\uparrow}_{i}\ket{\downarrow}_{i+1}+
q^{\half}\ket{\downarrow}_{i}\ket{\uparrow}_{i+1}}
\haak{q^{-\half}\bra{\uparrow}_{i}\bra{\downarrow}_{i+1}+
q^{\half}\bra{\downarrow}_{i}\bra{\uparrow}_{i+1}}.
\end{equation}
This expression is indeed compatible with the rules \eqref{eq:TLdef}.
This choice differs from the quantum group convention, in order to
avoid a complication in the even periodic system.
We can rewrite this in terms of the Pauli matrices as
\begin{equation}\label{hamloc}
e_{i}=\half\haak{\sigma^{x}_{i}\sigma^{x}_{i+1}+\sigma^{y}_{i}\sigma^{y}_{
i+1}-\half\sigma^{z}_{i}\sigma^{z}_{i+1}
-\frac{\imath\sqrt{3}}{2}\haak{\sigma^{z}_{i}-\sigma^{z}_{i+1}}+\half}.
\end{equation}

When reflecting boundary conditions are imposed the spin Hamiltonian follows from
\eqref{hamilr} and \eqref{hamloc}:
\begin{equation}\label{hamref}
H^{\text{ref}}=-\sum_{i=1}^{L-1}\half\haak{\sigma^{x}_{i}\sigma^{x}_{i+1}+\sigma^{y
}_{i}\sigma^{y}_{
i+1}-\half\sigma^{z}_{i}\sigma^{z}_{i+1}
-\frac{\imath\sqrt{3}}{2}\haak{\sigma^{z}_{i}-\sigma^{z}_{{i+1}}}-\frac{3}{2}}.
\end{equation}

For periodic boundary conditions we have to distinguish
between systems of odd and even lengths.
For even periodic system, it is possible for a loop to wind round
the cylinder while for odd periodic systems the presence of the unpaired line
will prevent this. Therefore, in case of an odd periodic system, the
corresponding XXZ Hamiltonian is obtained from \eqref{hamloc} and \eqref{hamilp}.
We thus obtain:
\begin{equation}\label{hamodd}
H^{\text{odd}}=-\sum_{i=1}^{L}\half\haak{\sigma^{x}_{i}\sigma^{x}_{i+1}+
\sigma^{y}_{i}\sigma^{y}_{i+1}-\half\sigma^{z}_{i}\sigma^{z}_{i+1}-\frac{3}{2}}.
\end{equation}
Here $\sigma_{0}=\sigma_{L}$.

For an even periodic system, we have to give a weight of 1 to loops that wind round
the cylinder. This can be done by introducing a seam at the
boundary between the sites $L$ and 1. We assign a weight of $q$ to an arrow
pointing from $L$ to 1, while an arrow pointing in the opposite direction
will be assigned a weight of $q^{-1}$. As a result $e_{L}$ is modified:
\begin{equation}
e_{L}=\haak{q^{-\frac{3}{2}}\ket{\uparrow}_{L}\ket{\downarrow}_{1}+
q^{\frac{3}{2}}\ket{\downarrow}_{L}\ket{\uparrow}_{1}}
\haak{q^{\half}\bra{\uparrow}_{L}\bra{\downarrow}_{1}+
q^{-\half}\bra{\downarrow}_{L}\bra{\uparrow}_{1}}.
\end{equation}
It is convenient to rewrite this in terms of $\sigma^{+}$ and
$\sigma^{-}$:
\begin{equation}
e_{L}=\sigma^{+}_{1}\sigma^{-}_{L}q^{2}+
\sigma^{-}_{1}\sigma^{+}_{L}q^{-2}-\half\haak{
\half\sigma^{z}_{1}\sigma^{z}_{L}+\frac{\imath\sqrt{3}}{2}\haak{
\sigma^{z}_{L}-\sigma^{z}_{1}}-\half}.
\end{equation}
The Hamiltonian is thus given as:
\begin{equation}\label{hameven}
\begin{split}
H^{\text{even}}=&-\rhaak{\sum_{i=1}^{L-1}\half\haak{\sigma^{x}_{i}\sigma^{x}_{
i+1}+\sigma^{y}_{i}
\sigma^{y}_{i+1}-\half\sigma^{z}_{i}\sigma^{z}_{i+1}}-\frac{3}{2}}\\
\mbox{}&-\sigma^{+}_{1}\sigma^{-}_{L}q^{2}-\sigma^{-}_{1}\sigma^{+}_{L}q^{-2}+\kwart
\sigma^{z}_{1}\sigma^{z}_{L}+\frac{3}{4}.
\end{split}
\end{equation}

To obtain the ground states of the Hamiltonians \eqref{hamref}, \eqref{hamodd} and \eqref{hameven} from the eigenvectors of the corresponding loop models, one has to assign orientations to all the half loops of the connectivity states contributing to the eigenvectors in all possible ways and assign the correct phase factors. A half loop starting at $i$ and ending at $j$ for $j>i$ corresponds to: 
\begin{equation}\label{convert}
q^{-1/2} \ket{\uparrow}_{i} \ket{\downarrow}_{j} + q^{1/2} \ket{\downarrow}_{i} \ket{\uparrow}_{j}.
\end{equation}
For an even periodic system one has to take into account the presence of the seam. A half loop starting at $i$ and ending at $j$ for $i>j$ that crosses the seam corresponds to the term:
\begin{equation}
q^{-3/2} \ket{\uparrow}_{i} \ket{\downarrow}_{j} + q^{3/2} \ket{\downarrow}_{i} \ket{\uparrow}_{j}.
\end{equation}
 
\subsection{Open and Mixed Boundary Conditions}
When open or mixed boundary conditions are imposed, we have to take into account the fact that
loops can end at the boundary. We have to give weights to arrows pointing in or
out of the left and the right boundary, such that lines starting from one 
boundary and ending on the same or the other boundary get a weight of 1. We denote these weights as $w^{d}_{v}$, where $d$ can be $l$ for the left boundary or $r$ for the right boundary, while
$v$ can be $-$ for an arrow pointing into the boundary or $+$ for an arrow
pointing out of the boundary.

Demanding that a line starting from the left boundary and ending on the left 
boundary
has a weight of 1 yields:
\begin{equation}\label{cond1}
w^{l}_{+}w^{l}_{-}\sqrt{3}=1.
\end{equation}
Similarly, demanding that a line starting from the right boundary and ending on
the right boundary
has a weight of 1 yields:
\begin{equation}\label{cond2}
w^{r}_{+}w^{r}_{-}\sqrt{3}=1.
\end{equation}
Demanding that a line starting on one boundary ending on the other boundary
has a weight of 1 yields
\begin{equation}\label{cond3}
w^{l}_{+}w^{r}_{-}+w^{l}_{-}w^{r}_{+}=1.
\end{equation}
The choice:
\begin{eqnarray}
w^{l}_{+}&=&a 3^{-\kwart}\\
w^{l}_{-}&=&a^{-1} 3^{-\kwart}\\
w^{r}_{+}&=&\frac{a}{2}\haak{3^{\kwart}+\imath 3^{-\kwart}}\\
w^{r}_{-}&=&\frac{1}{2a}\haak{3^{\kwart}-\imath 3^{-\kwart}}
\end{eqnarray}
satisfies the equations (\ref{cond1}), (\ref{cond2}) and (\ref{cond3}) for general $a$.

We can now use these weights to transform the operators $h_{0}$ and $h_{L}$
(see Fig.\ \ref{fig:hdef}) into spin operators. We obtain:
\begin{eqnarray}
h_{0}&=&\haak{q^{-\kwart}w^{l}_{-}\ket{\downarrow}_{1}+q^{\kwart}w^{l}_{+}\ket{\uparrow}_{1}}
\haak{q^{-\kwart}w^{l}_{+}\bra{\downarrow}_{1}+q^{\kwart}w^{l}_{-}\bra{\uparrow}_{1}}\\
h_{L}&=&\haak{q^{\kwart}w^{r}_{-}\ket{\downarrow}_{L}+q^{-\kwart}w^{r}_{+}\ket{\uparrow}_{L}}
\haak{q^{\kwart}w^{r}_{+}\bra{\downarrow}_{L}+q^{-\kwart}w^{r}_{-}\bra{\uparrow}_{L}},
\end{eqnarray}
acting in the state space of the first and the last spin respectively.
We can rewrite this in matrix form as:
\begin{equation}
h_{0}=\haak{
\begin{array}{cc}
w^{l}_{+}w^{l}_{-}q^{\half}&\haak{w^{l}_{+}}^{2}\\
\haak{w^{l}_{-}}^{2}&w^{l}_{+}w^{l}_{-}q^{-\half}\\
\end{array}
}_{1}
\end{equation}
\begin{equation}
h_{L}=\haak{
\begin{array}{cc}
w^{r}_{+}w^{r}_{-}q^{-\half}&\haak{w^{r}_{+}}^{2}\\
\haak{w^{r}_{-}}^{2}&w^{r}_{+}w^{r}_{-}q^{\half}
\end{array}
}_{L}.
\end{equation}

The Hamiltonian for mixed boundary conditions \eqref{hamilm} can thus be written as:
\begin{equation}
\begin{split}\label{hammixed}
H^{\text{mixed}}=&-\rhaak{\sum_{i=1}^{L-1}\half\haak{\sigma^{x}_{i}\sigma^{x}_{i+1}+\sigma^{y
}_{i}\sigma^{y}_{
i+1}-\half\sigma^{z}_{i}\sigma^{z}_{i+1}
-\frac{\imath\sqrt{3}}{2}\haak{\sigma^{z}_{i}-\sigma^{z}_{{i+1}}}-\frac{3}{2}}}\\
&\mbox{}+1-h_{0}.
\end{split}
\end{equation}
And the Hamiltonian for open boundaries \eqref{hamilo} becomes:
\begin{equation}
\begin{split}\label{hamopen}
H^{\text{open}}=&-\rhaak{\sum_{i=1}^{L-1}\half\haak{\sigma^{x}_{i}\sigma^{x}_{i+1}+\sigma^{y
}_{i}\sigma^{y}_{
i+1}-\half\sigma^{z}_{i}\sigma^{z}_{i+1}
-\frac{\imath\sqrt{3}}{2}\haak{\sigma^{z}_{i}-\sigma^{z}_{{i+1}}}-\frac{3}{2}}}\\
&\mbox{}+2-h_{0}-h_{L}.
\end{split}
\end{equation}

To obtain the ground states of the Hamiltonians \eqref{hammixed} and \eqref{hamopen} from the eigenvectors of the corresponding loop models, one proceeds in the same way as in the case of periodic and reflecting systems. A half loop starting at a point $i$ and ending at a point $j$ for $j>i$ produces the term \eqref{convert}. A quarter loop starting at the point $i$ and ending at the left boundary corresponds to the term:
\begin{equation}
q^{-\kwart}w^{l}_{-}\ket{\downarrow}_{i}+q^{\kwart}w^{l}_{+}\ket{\uparrow}_{i}.
\end{equation}
A quarter loop starting at the point $i$ and ending at the right boundary corresponds to:
\begin{equation}
q^{\kwart}w^{r}_{-}\ket{\downarrow}_{i}+q^{-\kwart}w^{r}_{+}\ket{\uparrow}_{i}.
\end{equation}
\section{Conclusion}
We have presented new conjectures for correlations in the dense O$(1)$ loop model on finite by infinite square lattices. We have obtained results for periodic, reflecting, open and mixed boundary conditions. The obtained correlations involve probabilities that points on a row are connected by loops via the half space below the row in certain ways. Also a conjecture has been obtained relating the ground state for a system with open identified boundaries to the FPL model on a rectangular grid.

\section{Acknowledgements}
This work was supported by Stichting FOM, which is part of the Dutch foundation of scientific research NWO, and also by the Australian Research Council. We thank Christian Krattenthaler for helping us to find the closed form expression of the determinant in \eqref{deter} and Pavel Pyatov for correcting an error regarding the largest component of the eigenvector for even open systems.

\appendix
\section{Appendix: The boundary term}

The Hamiltonian (\ref{hamilp}) for the periodic system commutes with a
whole family of transfer matrices for the percolation or dense O($n$=1)
problem. This is because the Boltzmann weights satisfy the Yang-Baxter
equation. The Hamiltonians with a boundary, open or reflecting, also
commute with a family of transfer matrices. In this case, besides the
Yang-Baxter equation we need the reflection equation\cite{Sklyanin}, which
should be satisfied by the boundary weights. In this appendix we study
the reflection equation for the dense O($n$) model.

To set the notation we consider the Yang-Baxter equation of in the form
\begin{equation} 
E_j(u) E_{j+1}(u+v) E_j(v) = E_{j+1}(v) E_j(u+v) E_{j+1}(u) \label{YB}
\end{equation}
where the operators $E_j(u)$ are a linear combination of the identity
and the monoid $e_j$, in a ratio specified by the spectral
parameter. The relations \eqref{eq:TLdef} among the $e_j$ are
sufficient to show that (\ref{YB}) is solved by
\begin{equation} E_j(u) = \sin(\lambda - u) + e_j \sin u \end{equation}
This operator suffices to construct the transfer matrix for the periodic
system.

The operators $E$ are also an ingredient to the family of commuting
transfer matrices for a system with a boundary. In this case the
transfer matrix involves two consecutive rows of vertices
\cite{doublerow} of which the horizontal legs are joint at the boundary
by a boundary weight, or K-matrix.  The left boundary operator $F_0(u)$
must satisfy the reflection equation\cite{Sklyanin}:
\begin{equation} \label{LRE}
F_0(u) E_1(u+v) F_0(v) E_1(v-u) = E_1(v-u) F_0(v) E_1(u+v) F_0(u),
\end{equation}
where now $F_0(u)$ is a linear combination of the identity and the
operator $h_0$, which satisfies \eqref{relhe}.
The coefficient in $F_0$ of the identity is the weight with which the
two legs are simply connected, as in fig.\ \ref{fig:refl}. The coefficient of
$h_0$ is the weight with which both legs are connected to the boundary.
The solution to this equation is given by
\begin{equation}\label{Kmatrix}
F_0(u) = A + B \sin(\lambda/2-2 u) + 2 B \cos(\lambda/2) \sin(2u) h_0 .
\end{equation}
for general $\lambda$, and arbitrary constants $A$ and $B$.
This equation is the central result of this appendix.
For the right boundary we have the analogous equation
\begin{equation} \label{RRE}
F_L(u) E_{L-1}(u+v) F_L(v) E_{L-1}(v-u) = 
E_{L-1}(v-u) F_L(v) E_{L-1}(u+v) F_0(u),
\end{equation}

The trivial solution of (\ref{LRE}) is $B=0$, and say $A=1$, 
in which case $F_0$ is simply the
identity. This corresponds with reflecting boundary conditions, in which
the horizontal lines of the double row transfer matrix are always 
mutually connected. Since there is no $u$-dependence, the Hamiltonian
for this case (\ref{hamilr}) does not have a boundary term.

When $B$ is non-zero the logarithmic derivative of the transfer matrix
will have a term proportional with $h_0$. The coefficient of this term
is 
\[
4 B \cos (\lambda/2)/\left[ A + B\sin(\lambda/2) \right].
\]
The coefficients of the terms $e_i$ are equal to $2/\sin\lambda$.
For $\lambda=\pi/3$, the coefficients of $h_0$ and $e_i$ 
are equal when $A=B$. This is the case in
the expressions (\ref{hamilo}) and (\ref{hamilm}).

Finally, it is of interest to consider the nature of the boundary of the
transfer matrix in the point $u=\lambda/2$ in which the transfer matrix
has left-right reflection symmetry.
In the case of reflecting boundaries, $B=0$, 
the horizontal lines at the boundary of the even rows
are simply connected to that of the odd row below it, as in
fig.\ \ref{fig:refl}. 
For open boundary conditions we choose $A$ and $B$ equal as indicated above.
The common magnitude of $A$ and $B$ is immaterial and simply
multiplies the entire transfer matrix. In the convenient choice  
$A=B=1/2$ at $\lambda=\pi/3$ the boundary weight can be interpreted as a
probability distribution: the coefficient in (\ref{Kmatrix}) of $h_0$ and
the identity are $3/4$ and $1/4$ respectively, to be interpreted as the
probability that the horizontal legs of two consecutive rows are
connected to the boundary, or to each other, respectively.

\end{document}